\documentclass[lettersize,journal]{IEEEtran}
\usepackage{amsmath,amsfonts}
\usepackage{algorithm}
\usepackage{array}
\usepackage{textcomp}
\usepackage{stfloats}
\usepackage{url}
\usepackage{verbatim}
\usepackage{graphicx}
\usepackage{cite}

\usepackage{algpseudocode}
\usepackage{amsmath,amssymb,amsfonts}
\usepackage{booktabs}
\usepackage{calc}
\usepackage{filecontents}
\usepackage{float}
\usepackage{mathtools}
\usepackage{multirow}
\usepackage{pgfplots}
\pgfplotsset{compat=1.15}
\usepackage{pgfplotstable}
\usepackage{ragged2e}
\usepackage{subcaption}
\usepackage{adjustbox}

\begin{document}

\title{Reconstruction-based LSTM-Autoencoder for Anomaly-based DDoS Attack Detection over Multivariate Time-Series Data}

\author{Yuanyuan Wei, Julian Jang-Jaccard, Fariza Sabrina, ~\IEEEmembership{Member,~IEEE}, Wen Xu, \\Seyit Camtepe, ~\IEEEmembership{Senior Member,~IEEE}, and Aeryn Dunmore
\thanks{Yuanyuan Wei is with the CybersecurityLab, Comp Sci/Info Tech, Massey University, Auckland, 0632, NEW ZEALAND (e-mail: y.wei1@massey.ac.nz).}
\thanks{Julian Jang-Jaccard is with the CybersecurityLab, Comp Sci/Info Tech, Massey University, Auckland, 0632, NEW ZEALAND (e-mail: j.jang-jaccard@massey.ac.nz).}
\thanks{Fariza Sabrina is with the School of Engineering and Technology, Central Queensland University, Sydney NSW 2000, AUSTRALIA (e-mail: f.sabrina@cqu.edu.au).}
\thanks{Wen Xu is with the CybersecurityLab, Comp Sci/Info Tech, Massey University, Auckland, 0632, NEW ZEALAND (e-mail: w.xu2@massey.ac.nz).}
\thanks{Seyit Camtepe is with the CSIRO Data61, AUSTRALIA (e-mail: Seyit.Camtepe@data61.csiro.au).}
\thanks{Aeryn Dunmore is with the CybersecurityLab, Comp Sci/Info Tech, Massey University, Auckland, 0632, NEW ZEALAND (e-mail: a.dunmore@massey.ac.nz).}

\thanks{Manuscript received April 19, 2021; revised August 16, 2021.}}

\markboth{Journal of \LaTeX\ Class Files,~Vol.~14, No.~8, August~2021}%
{Shell \MakeLowercase{\textit{et al.}}: A Sample Article Using IEEEtran.cls for IEEE Journals}


\maketitle

\begin{abstract}
A Distributed Denial-of-service (DDoS) attack is a malicious attempt to disrupt the regular traffic of a targeted server, service, or network by sending a flood of traffic to overwhelm the target or its surrounding infrastructure. As technology improves, new attacks have been developed by hackers. Traditional statistical and shallow machine learning techniques can detect superficial anomalies based on shallow data and feature selection, however, these approaches can not detect unseen DDoS attacks. In this context, we propose a reconstruction-based anomaly detection model named LSTM-Autoencoder (LSTM-AE) which combines two deep learning-based models for detecting DDoS attack anomalies. The proposed structure of long short-term memory (LSTM) networks provides units that work with each other to learn the long short-term correlation of data within a time series sequence. Autoencoders are used to identify the optimal threshold based on the reconstruction error rates evaluated on each sample across all time-series sequences.  As such, a combination model LSTM-AE can not only learn delicate sub-pattern differences in attacks and benign traffic flows but also minimize reconstructed benign traffic to obtain a lower range reconstruction error, with attacks presenting a larger reconstruction error. In this research, we trained and evaluated our proposed LSTM-AE model on reflection-based DDoS attacks (DNS, LDAP, and SNMP). The results of our experiments demonstrate that our method performs better than other state-of-the-art methods, especially for LDAP attacks, with an accuracy of over 99\%.
\end{abstract}


\section{Introduction}
\IEEEPARstart{N}{etwork} traffic is increasing rapidly with the continued development of information and communication technology (ICT) due to advanced innovative technologies, including cloud computing, and big data. However, the rapid proliferation of innovative technologies and communication infrastructure brings the potential for cyberattacks and other threats to Internet users. In the area of cyber security attacks, one of the most dangerous threats is a distributed denial-of-service (DDoS) attack~\cite{haider2020deep,sahoo2019toward,yuan2017deepdefense}.

A DDoS attack is a form of network attack that attempts to overwhelm online services, websites, and web applications with malicious traffic from multiple compromised computer systems. It can also make simultaneous requests to the target server in order to exhaust the network resources and thereby deny normal online service to legitimate users or computer systems~\cite{sriram2020network,doriguzzi2020lucid,saied2016detection}. DDoS attacks are not only conducted against online services, web applications, and information infrastructure to cause downtime, but also to prevent legitimate users from purchasing products and using online services - such as emails, websites, and applications - and affecting program performance~\cite{saied2016detection}. As a result of the COVID-19 lockdown in 2020, there has been an increase in attacks on education, online shopping, and office work, as a large number of people are now studying, working, and shopping online, giving hackers greater opportunities~\cite{Alethea2021Azure}. In~\cite{microsoft2021microsoft} Azure Networking found there was a 25\% increase in DDoS attacks in the first six months of 2021 when compared with the fourth quarter of 2020. Moreover, Azure mitigated approximately 35 thousand attacks against its global infrastructure in the last six months of 2021, which increased from 43\% compared with the first six months of 2021. A white paper from Cisco~\cite{cisco2020cisco} predicted that nearly 300 billion mobile applications would be downloaded by 2023 and that DDoS attacks would rise to 15.4 million globally by 2023.


DDoS detection is becoming an urgent need because of the sophistication and diversification of attacks. For instance, difficult-to-track attackers and unknown or new attack types occur continuously, for example, zero-day attacks~\cite{yeom2022lstm,abdallah2021hybrid}. Detecting DDoS attacks becomes more and more difficult not only because a large proportion of attack traffic is similar to legitimate traffic, but also because of newer hybrid attack methods~\cite{yuan2017deepdefense,doriguzzi2020lucid}. Therefore, the detection and mitigation of DDoS attacks not only protects the network for legitimate users but also reduces financial loss for businesses~\cite{salahuddin2021chronos}. In order to detect and mitigate DDoS attacks, statistical techniques have been proposed in ~\cite{mirkovic2002attacking} to identify DDoS attacks. Furthermore, some machine learning approaches for signature, threshold, and statistics-based measurements have been proposed to distinguish DDoS attack traffic~\cite{alzahrani2018detection,vinayakumar2019deep}. However, traditional statistical and machine learning can not detect previously unseen DDoS attacks~\cite{doriguzzi2020lucid}. Moreover, most traditional statistical and machine learning-based detection approaches require better-selected features or defined thresholds~\cite{doriguzzi2020lucid,yuan2017deepdefense}. In contrast to traditional detection techniques, deep learning-based DDoS attack detection - such as Convolutional Neural Networks (CNNs)~\cite{doriguzzi2020lucid}, Recurrent Neural Networks (RNNs)\cite{yuan2017deepdefense}, Autoencoders~\cite{salahuddin2021chronos}, and so forth -  can offer better detection rates for DDoS network traffic~\cite{yuan2017deepdefense}. However, some limitations in existing deep learning-based detection need to be addressed, for example, Autoencoder models are sensitive to the anomalies in the training stage, and RNNs can better address historical sequence data, but face the shortcoming of the vanishing gradient problem. To address these issues, we propose a reconstruction-based hybrid deep learning model that combines the capabilities of long short-term memory (LSTM) and Autoencoders (AE) for detecting DDoS attacks, using the state-of-the-art CICDDoS2019 datasets. 

In this research, the LSTM model aims to solve the time-series sequence problem of DDoS traffic flow, while the AE is used to calculate the reconstruction loss in order to define the threshold and detect DDoS attacks. In order to overcome the shortcoming caused by sensitivity to anomalies in the training process, we use only benign traffic from the DDoS dataset to train our model and minimize the reconstruction error. Furthermore, the LSTM can learn the time series sequence of DDoS network traffic continuously but learns the delicate difference between attacks and benign traffics based on the time window length section. A combination model LSTM-AE can learn delicate differences in sub-patterns between attacks and benign traffic while minimizing the reconstructed benign traffic to obtain a lower range reconstruction error. Our experimental results showed that the proposed LSTM-AE model achieves better performance in processing reconstruction-based time-series data than other comparable proposed models. The main contributions of our proposed model are as follows:

\textbf{Summary of Original Contributions}

\begin{itemize}
    \item We propose a novel time-series anomaly detection architecture that leverages reconstruction-based LSTM-AE for efficient DDoS attack detection. In our proposed model, the LSTM networks are comprised of multiple LSTM units that work with each other to learn the long short-term correlation of data within a time series sequence. An autoencoder is used to identify the optimal threshold based on the reconstruction error rates evaluated across all time-series sequences. This threshold is used to identify anomalies.
	\item We apply our proposed LSTM-AE model against the reflection-based DDoS attacks - DNS, LDAP and SNMP. The model is trained on normal time-based traffic flow features using a subset of traffic flow information over a fixed-time window length.
	\item A novel anomaly score technique is proposed to calculate the MAE value of each traffic flow, which can be calculated flexibly based on different fixed-time window lengths.
	\item We performed tests on the state-of-the-art CICDDoS2019 dataset, and compared the performance of our proposed model with other similar approaches that use different aspects of LSTM and/or AE. Our experimental results, based on the comprehensive set of evaluation criteria, demonstrate that our proposed model can effectively detect anomalies reaching a detection accuracy exceeding 99\%.
\end{itemize}

The rest of this paper is structured as follows: Section~\ref{sec:rw} introduces related works in the field of DDoS attack detection.  Section~\ref{sec:method} introduces our methodology. Section~\ref{sec:re} illustrates the experimental setup and  Section \ref{sec:Experiments} details the analysis of our results evaluated on the various reflection-based attack types, including DNS, LDAP, and SNMP datasets. Section~\ref{sec:Conclusion} concludes the paper with the planned future works.

\section{Related Work}\label{sec:rw}
In recent years, anomaly detection has attracted extensive attention in literature exploring machine learning techniques. In this paper, we review the issues of variable length of DDoS anomaly detection, areas closely related to our contributions.

Sharafaldin et al.~\cite{sharafaldin2019developing} generated a dataset titled CICDDoS2019 and classified benign traffic and attacks based on 4 machine learning methods, including ID3, Random Forest, Naïve Bayes, and Multinomial Logistic Regression. The evaluation result shows the highest accuracy from RF and ID3. Jia et al.~\cite{jia2020flowguard} proposed an IoT DDoS defense technique named FlowGuard, and constructed LSTM and CNN techniques for DDoS identification and classification based on simulated data and CICDDoS2019 dataset to identify, classify, and mitigate DDoS attacks. They used a CNN to better classify all malicious flows, then employed the LSTM technique as an identification module. This was not only able to capture significant features of flows to identify benign samples but also to apply a softmax function on the output layer to distinguish between benign and malicious traffic. The above researches used flow-based statistical features, which were extracted from CICDDoS2019 dataset for DDoS attack detection.

Novaes et al.~\cite{novaes2020long} introduced two scenarios for detecting anomalies and mitigating attacks on Software-Defined Networks (SDNs) using LSTM-FUZZY techniques. Firstly, the authors used an LSTM network semi-supervised learning technique to predict benign univariate time series behaviour of IP flows, followed by classifying attacks with a Fuzzy logic technique. There were two datasets used in this research, including the SDN dataset and the CICDDoS2019 dataset. Aydin et al.~\cite{aydin2022long} proposed an LSTM-based system (LSTM-CLOUD) to detect and prevent DDoS attacks in a public cloud network environment through experimentation on CICDDoS2019 dataset. The authors built the LSTM models with two hidden layers, three dense and dropout layers, and used the sigmoid function to classify benign and DDoS attacks (anomalies). The model performed to a high accuracy rate of 99.83\%, but this work only classified 3 attack types of attacks - UDP, MSSQL, SYN - as well as benign traffic samples. 

Nezhad et al.~\cite{nezhad2016novel} normalized two features (packets and source IP addresses) on a 1-minute time series interval by using a Box-Cox transformation. They then used a statistical time series analysis technique called ARIMA to predict the number of packets. 

Ergan et al.~\cite{ergen2019unsupervised} introduced LSTM-based neural networks to detect anomalies in a time series in an unsupervised manner. The author employs an LSTM-based network technique to obtain the fixed-length sequence data. They utilized the OC-SCM and SVDD algorithms together with a scoring function to detect anomalies.

Salahuddin et al.~\cite{salahuddin2021chronos} proposed a time-based Autoencoder technique named Chronos to detect DDoS traffic anomalies with aggregating features. One of the contributions in this research was the threshold they selected to highlight the efficiency of their anomaly detection system. They utilised threshold selection heuristic maximizes the F1 score. To detect anomalies effectively, they implemented various window sizes on different DDoS attacks. As a result, the proposed Chronos system achieves an F1-score of 99\% for most attack types and over 95.86\%  for all attack type performance measurements by using two-time windows along with the selected heuristic threshold.

Fouladi et al.~\cite{fouladi2019anomaly} used the CAIDA dataset to train and evaluate the K-SVD and BMP (Basic Matching Pursuit) algorithms and a SOM (Self-Organizing Map) model, using sparse coding and frequency domain for DDoS attack anomaly detection. They extracted normal time series data using a K-SVD algorithm and applied a BMP method to train and evaluate the benign data to estimate the sparse coefficients. They then used the normal data on the SOM model to obtain a SOM lattice. This enabled them to calculate the minimum Euclidean Distance between corresponding coefficients and use the SOM lattice to distinguish between normal and attack behaviors.

Although many researchers above have introduced statistical and machine learning techniques to detect anomalies in the DDoS datasets, it was also a challenging task worthy of further study, especially for distinguishing subtle differences in benign and attack traffics. In this research, we propose a hybrid deep learning reconstruction-based LSTM-Autoencoder model for anomaly-based DDoS attack detection. Our model uses a specific selection of features over multivariate time-series data analysis. Compared to the above statistical and machine learning-based studies, the differences in this research can be detailed as follows: 1) Our LSTM-AE neural network model conducts training and testing in an unsupervised manner (without labels), this is combined with the reconstruction error used to detect anomalies with several different time window lengths; 2) Anomaly detection in our proposed model gave experimental results for that DDoS attack anomaly detection with high accuracy on the CICDDoS2019 time-series dataset; 3) This experiment performed better in terms of the performance matrix - including precision, recall, and F1-score - than benchmark studies on the same CICDDoS2019 dataset, such as \cite{sharafaldin2019developing}, which used machine learning techniques (including Random forest, Naive Bayes, etc.) to detect DDoS attacks. 

\begin{figure*}[t]
	\centering
	\includegraphics[width=0.7\linewidth]{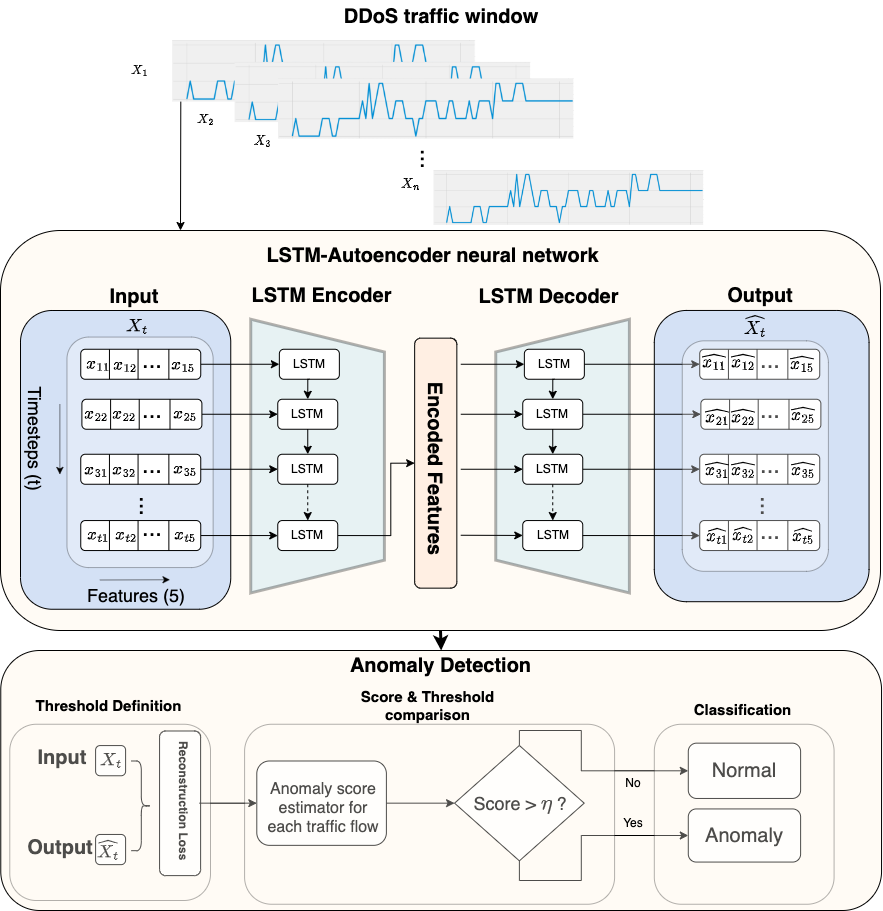}
	\caption{Overview of our proposed model}
	\label{alg:LSTM_AE}
\end{figure*}

\section{LSTM-Autoencoder Anomaly Detection}
\label{sec:method}
\subsection{Overview of Our Framework}
This section provides an overview of our reconstruction-based time-series anomaly detection system.
One of the DDoS traffic characteristics is collecting correlated temporal sequences, and the deep learning technique of LSTM is a good algorithm to deal with this temporal problem. Thus, we adopted the LSTM model for the LSTM-Autoencoder neural network for the Encoder and Decoder stages as its mechanism better captures DDoS flow information by feeding each flow at each time step. We propose a hybrid machine learning model which combines an LSTM neural network and an autoencoder. We apply our model to the multivariate time-series dataset CICDDoS2019. 
As illustrated in Fig.~\ref{alg:LSTM_AE}, our LSTM-AE model builds the LSTM networks on the encoder and decoder schemes of an Autoencoder. The encoder obtains the sequence of the high-dimensional input data as a fixed-size vector. Using the memory cells of LSTM, the data processed by the encoder scheme retains the dependencies across multiple data points within a time-series sequence. This stage reduces the high-dimensional input vector representation into a low-dimensional representation. The decoder-LSTM reproduces the fixed-size input sequence from the reduced representation of the input data in the latent space, while reconstruction error rates determine the classification threshold.  Fig.~\ref{alg:LSTM_AE} depicts the operation of a model for DDoS anomaly detection. Phase 1 is data pre-processing based on the selected characterization of each traffic flow fed into the LSTM-AE neural network. Phase 2 is the training and testing process, in which the threshold is obtained based on minimizing the reconstruction error on benign traffic. The DDoS anomaly detection is the last phase in the diagram, which determines anomaly scores by calculating the reconstruction error of each traffic flow between the time-series sequence's original input and the reconstructed output.


\subsection{Feature usage and Input Sequence Data}

Fig~\ref{alg:LSTM_AE} shows an overview of our proposed LSTM-AE model. To obtain the DDoS traffic flow data, we used the public dataset CICDDoS2019~\cite{sharafaldin2019developing} from the Canadian Institute for Cybersecurity. In \cite{sharafaldin2019developing}, the extracted network traffic features are stored in a CSV file, and each CSV file includes different DDoS attack types (anomalies) and benign traffic. In this research, we used three reflection-based DDoS attack types to detect anomalies, specifically DNS, SNMP and LDAP. As shown in the first part of the DDoS traffic window in Fig~\ref{alg:LSTM_AE}, all DDoS traffic flows are separated into $n$ subsets based on a selected window length. 

As our LSTM-AE model for detecting anomalies depends on learning the network traffic patterns, it is important to use high correlation features to obtain high performance in measurements such as accuracy. Using the research on the state-of-the-art CICDDoS2019 DDoS attack dataset in~\cite{sharafaldin2019developing}, the author specifies the importance of selecting the top-n features which correspond to the weight and mean value of different attack types. In this research, we have also used these top-n features as depicted in \cite{sharafaldin2019developing} from the CICDDoS2019 dataset. As such 'Max Packet Length', 'Fwd Packet Length Max', 'Fwd Packet Length Min', 'Average Packet Size', 'Min Packet Length' is used as selected features for our multivariate time-series anomaly detection. After feature selection, the input data has to be reshaped into a 2-dimensional vectors before the data is fed into the required LSTM encoder input layer. The original DDoS dataset is comprised of a series of time sequence $[X_1, X_2, X_3,...,X_n]$. Each sequence $X$ with a fixed T-length time window data $[x_1, x_2, x_3,...,x_t]$ is created where $x_t \in R{^m}$ represents an $m$-features input at time-instance $t$. They are then reshaped into 2-D (2-dimensional) arrays, representing samples and time steps. For example, a sequence of the DDoS attack data is converted into a 2-D array where each dimension indicates the list of samples at 10 time steps with $n$ features.


\subsection{LSTM-AE Model Architecture}
\subsubsection{Long Short Term Memory}
\begin{figure}[h]%
    \centering
    \subfloat[\centering LSTM]{{\includegraphics[width=4.6cm]{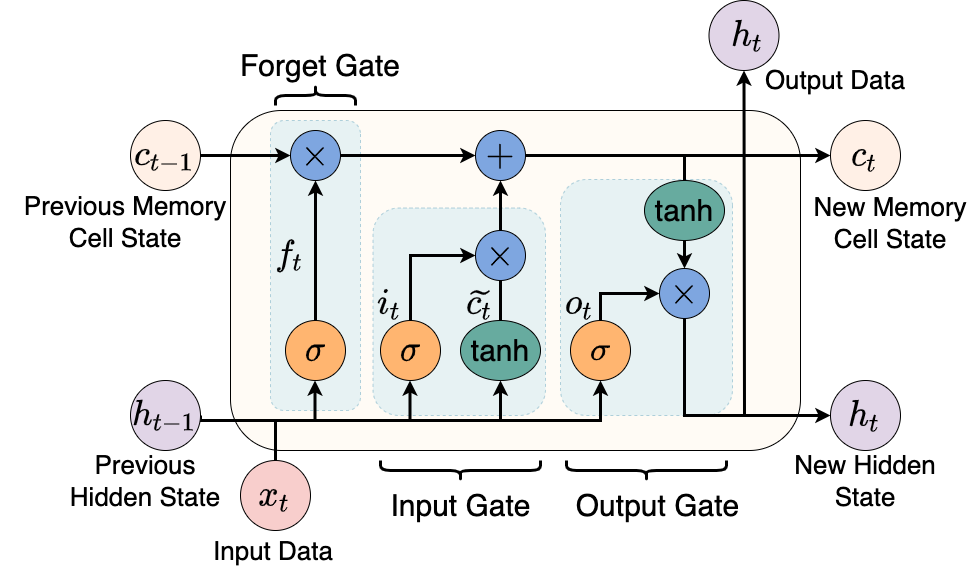} }
    \label{fig:LSTM}}
    \subfloat[\centering Autoencoder]{{\includegraphics[width=3.1cm]{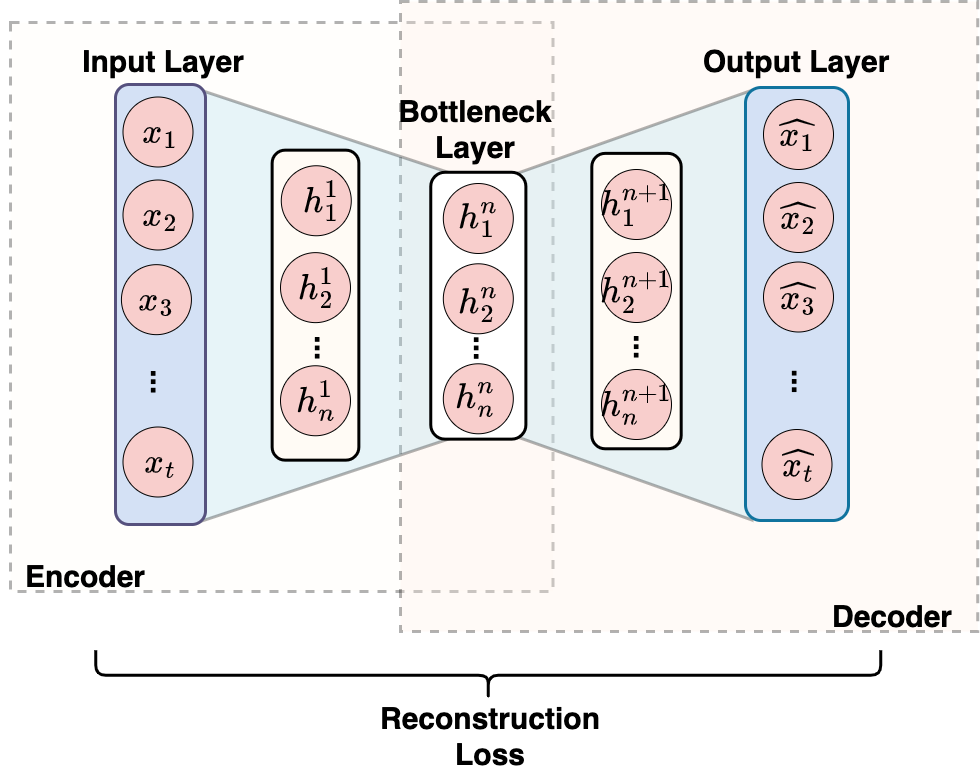} }
    \label{fig:AE}}%
    \caption{Neural Networks architectures: (a) LSTM; (b) Autoencodre.}%
    \label{fig:architectures}%
\end{figure}

The LSTM network is a variation of a Recurrent Neural Network (RNN) that addresses the gradient vanishing and exploding problems of RNNs and can process long term sequences between data samples at any given time from many history time steps. The architecture of the LSTM network is suitable for processing time series sequence data and provides the capability of forgetting the historical data from each memory cell before updating the memory cell with new data. As illustrated in Fig.~\ref{fig:architectures}(a), the LSTM network contains memory cells $c_{t}$ and multiple gate units, including the forget gate $f_{t}$, the input gate $i_{t}$ and the output gate $o_{t}$. These three gate units control the state of memory cells. For example, at time step $t$, the forget gate decides which bits of the cell state are useful given both the previous hidden state and new input data. The LSTM network can omit insignificant information (values) from the cell state of the forget gate and can recognize significant information (values) to keep and update in the cell state. Using this network architecture, the LSTM model performs to a high standard when capturing long-term patterns in time series sequence data. 




\subsubsection{Autoencoder}
An Autoencoder (AE) is an unsupervised neural network model that is not only used for feature selection and dimension reduction but also can be used for reconstruction-based Encoder-Decoders to detect anomalies. The typical architecture of an autoencoder is composed three components: an input layer, one or more hidden layer(s), and an output layer. The operations of an autoencoder for detecting anomalies can be divided into Encoding, Decoding, and Reconstruction Loss as illustrated in Fig. \ref{fig:architectures}(b). The encoder compresses the high dimension input data and maps it to low dimensional representations $h$, in the bottleneck layer while the decoder decompresses the encoded representation and reconstructs to the output $\hat{x}$. Typically, the autoencoder is trained by utilizing the MAE equation as a loss function to minimize the reconstruction error between the output $\hat{x_t}$ and input $x_t$.



\subsubsection{LSTM-Autoencoder}
We have addressed the issue of anomaly detection in capturing normal phenomena through time-based traffic flow from the DDoS attacks dataset. An LSTM-AE model need not only be applied to address the problem of feed-forward neural networks but also may be utilized to learn patterns in time-based sequence data, making them suitable for time-series anomaly detection \cite{said2020network, yin2017deep}. Our proposed LSTM-AE model includes an autoencoder that utilizes the LSTM network as a hidden layer in both the Encoder and Decoder schemes. This is followed by an Encoded Features layer and an output layer respectively, which calculate the reconstruction error to detect anomalies. The role of the LSTM network in our proposed scheme is to learn the patterns of data based on time-series DDoS signals. We combine this with an autoencoder to learn the best encoder-decoder scheme to detect anomalies. The output of the LSTM Encoder and Decoder are then compared with the original input data and the reconstruction error is backpropagated through the architecture to update the weights of the neural network.

Our LSTM-AE model employs an autoencoder that utilizes the LSTM network as a hidden layer in both Encoder and Decoder schemes followed by an Encoded Features layer and an output layer respectively. These calculate the reconstruction error to detect anomalies. The role of the LSTM network in our proposed scheme is to learn the patterns of data based on selected time window length sequences. Our proposed LSTM-AE is composed of six layers - an Input, an LSTM Encoder, an LSTM Decoder, a RepeatVector, a TimeDistributed and an Output layer.
The LSTM Encoder obtains the sequence of high-dimensional input data as a fixed-size vector and compresses the input data into a low-dimensional hidden representation. Within an LSTM cell, for an input time series sequence data $X_1$ = [$x_1, x_2, ..., x_t$] where $t$ represents the time steps, each $x_t$ calculation is performed using the following: 

 \begin{equation}\label{eq:ft}
 	{f_t} = \sigma ({w_f}[H_{t-1},x_t] + {b_f})
 \end{equation}
 \begin{equation}\label{eq:it}
 	{i_t} = \sigma ({w_i}[H_{t-1},x_t] + {b_i})
 \end{equation}
 \begin{equation}\label{eq:tildeCt}
 	{\tilde{C_t}} = tanh ({w_c}[H_{t-1},x_t] + {b_c})
 \end{equation}
 \begin{equation}\label{eq:Ct}
 	{C_t} = {f_t} \odot {C{_{t-1}}} + {{i_t} \odot {\tilde{C_t}}}
 \end{equation}
 \begin{equation}\label{eq:ot}
 	{o_t} = \sigma ({w_o}[H_{t-1},x_t] + {b_o})
 \end{equation}
 \begin{equation}\label{eq:ht}
 	{H_t} = {o_t} \odot tanh({C_t})
 \end{equation}

where 
 \begin{itemize}
 	\item $f_t$ represents the forget gate, $i_t$, $\tilde{C_t}$ and $C_t$ represents the input gate, and $o_t$ and $H_t$ represents the output gate.
    \item $w$ and $b$ are the weights and the bias of the forget gate, input gate and output gate.
    \item $H_{t-1}$ and $x_t$ present the concatenation of the hidden state and the current input respectively. 
    \item $\sigma$ is the activation function of each gate, and it outputs numbers in the range of [0, 1].
    \item $\odot$ represents element-wise multiplication.
 \end{itemize}

Next, after compressing input data into low-dimensional representation until it reaches the latent space (encoded features), all data presentation can be repeated $t$ times on the RepeatVector layer to feed into the LSTM Decoder layer. In this LSTM Decoder layer, the decoder scheme uses the same number of features (equal to the encoder features on the LSTM Encoder layer) to map the latent space representation back to a high-dimensional representation. We add a TimeDistributed layer in order to generate the output of the LSTM Decoder in time sequence. 

\textbf{Encoder:}
To illustrate the LSTM encoder stage, if the time step is set to 10 - seen LSTM cells working theory at the time step $t$ (show in Fig~\ref{fig:LSTM}) - the input includes the previous output of the hidden state ($h_{t_1}$) and the cell state ($C_{t_1}$), and the current input $x{_t}$ while output includes the new hidden state ($h_t$) and the new cell state ($C{_t}$), and an output $y{_t}$. At the time step $t+1$, the new input becomes a new input $x_{t+1}$ for the next set of cells, and the output is obtained from the last hidden state and cell state of the LSTM cell. This then becomes the new input of hidden state ($h_t$) and the cell state ($C_t$) information. Note that we discard the output ($y{_t}$) of each LSTM cell at each time step $t$ in the Encoder and preserve the initial state which is from the hidden state and cell state. Finally, the output of the last time step (10) is the hidden state ($h_{10}$) and the cell state ($c_{10}$), which is a summary of the entire 10 time steps. Therefore, the input vector of LSTM encoder is $10\times5$, where 10 is the time steps (or time window length) and 5 is the number of features. If we set the number of input features (units) of LSTM to 16, after passing through the LSTM Encoder layer the size of the output vector is $1\times16$.

Our LSTM Encoder architecture (see Fig.~\ref{alg:LSTM_AE}: LSTM-Autoencoder neural network) is defined as follows:
\begin{itemize}
    \item Each traffic flow in the Input layer has been reshaped into a 2D matrix. Note that each traffic flow is represented by a matrix $n \times t$, where $n$ represents the traffic flows, and $t$ represents the time steps. This form of a series of input data is able to capture DDoS traffic patterns based on the sliding window length. 
    \item Each LSTM Encoder layer consists of 16 LSTM units with "tanh" activation functions.
    \item We added a dropout layer (0.2) to prevent over-fitting in the Encoder part.
    \item The encoded features have lower dimensions than the number of input features.
\end{itemize}

\textbf{Decoder:}
Between the Encoder and Decoder layer, we added a RepeatVector Layer to create the copies of the $1\times16$ vector equal to the number of time steps, which we called Encoded Features. For example, the size of time steps in our model is 10, therefore this layer will create 10 copies of the encoded features as a two-dimensional vector $10\times16$. The output of this layer becomes the input of the LSTM Decoder at each time step, which means each copy of the encoded features at each time step will be the input of the next set of LSTM cells. As depicted, the difference of LSTM network between the Encoder and Decoder is that the output of each LSTM cell at each time step ($y{_t}$) in the Decoder cannot be discarded and are outputted as $y{_t}$. There are two reasons to get the output of each of the LSTM cells: firstly, for the added layer, TimeDistributed, the input can be a 3-dimensional vector, which means the output from the LSTM cells has to be a 3-dimensional vector; and second, to ensure the output of each time step is as close as possible to the input. Therefore, the LSTM Decoder representations obtained in low-dimensionality encoding are used as input in the decoder and there are utilised to reproduce the original input data using the LSTM network. This means the output from the last set of LSTM cells (copied $t$ times) then becomes the input to the LSTM Decoder network.

Each $1\times16$ set is fed as an input to the decoder which creates a 3 Layer network with 10 LSTM cell units. Each LSTM cell unit processes each $1\times16$ encoded feature. These LSTM units produce an output that represents the result of the learning from the encoded feature where the output is multiplied with the $1\times16$ vector created by the additional TimeDistribution layer. At the same time, each LSTM cell unit produces a second output containing the state of the data processed by the current LSTM cell which is passed to the next LSTM - with the exception of the last LSTM unit. The matrix multiplication between the output of each LSTM layer ($L$) ($10\times16$) and the TimeDistributed layer ($16\times1$) results in a vector with of size $10\times1$ - the same as the size of the input.

The LSTM Decoder (see Fig.~\ref{alg:LSTM_AE}: LSTM-Autoencoder Neural Network) architecture is defined as follows:
\begin{itemize}
    \item The encoded feature in the bottleneck layer will be the input of the LSTM Decoder.
    \item An LSTM Decoder layer consists of 16 LSTM units with "tanh" activation functions.
    \item We added a dropout layer (0.2) to prevent overfitting in the Decoder section.
    \item Typically, the output of LSTM Decoder has two outputs, including the output data (traffic flows) ($O{_t}$) and the new hidden state ($H{_t}$). The output of $H{_t}$ can be discarded. Thus, the output from the decoder part is the reconstructed feature of the same size and dimension as the input data.
\end{itemize}






The final aim of LSTM-AE is to reconstruct the input from the output, i.e. $\hat{X_1}$ $\approx$ $X_1$ where $X_1$ indicates the input while $\hat{X_1}$ indicates the output. 

\begin{figure}[htp]
	
	\subfloat[Training Phase]{%
		\includegraphics[clip,width=\columnwidth]{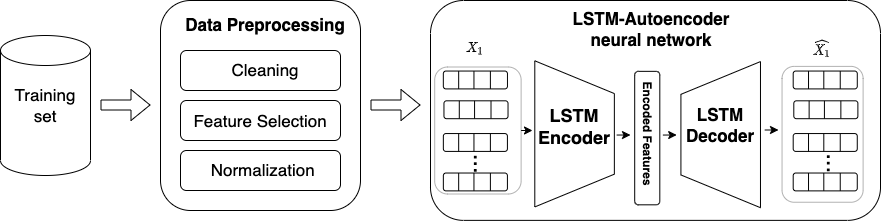}%
		\label{fig:training}
	}
	
	\subfloat[Testing Phase]{%
		\includegraphics[clip,width=\columnwidth]{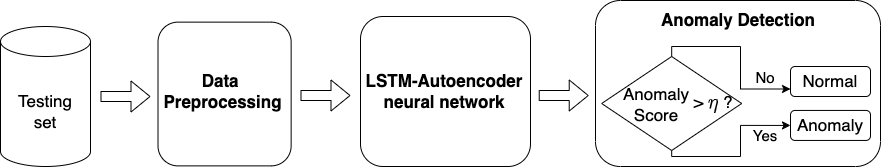}%
		\label{fig:testing}
	}
	
	\caption{Training and Testing Phase}
	\label{fig:training_testing}
\end{figure}

\textbf{LSTM-AE Training:} 
We divide the CICDDoS2019 dataset into a training set (70\%), a validation set (10\%), and a testing set (20\%). Note that the training and validation set consists of benign samples only, in order to train the proposed LSTM-AE model. The testing set consists of both benign and attack samples for detecting anomalies. The architecture of the proposed LSTM-AE is designed to minimize the reconstruction error between the original input and the reconstructed output based on time series sequence traffic flows. This LSTM model aims to learn the patterns in the data. Our motivation for using the Encoder and Decoder scheme to detect anomalies is the fact that their working scheme is able to detect anomalies based on the low distribution of normal data. As such, our training data set consists of only normal sequence data and this is utilized for training the proposed LSTM-AE model. The LSTM-AE is taught normal traffic behavior, using the benign samples. Our proposed LSTM-AE is an Encoder-Decoder unsupervised learning model, and no label is provided in the training phase. The training process is depicted in Fig.~\ref{fig:training_testing}(a). To address the problem of overfitting, dropout was added in both the encoder and decoder stages and set to 0.2. For the other settings in our model,  "Adam" is the optimization method, "tanh" is used as the activation function, and "MAE" can be chosen as a loss function.

\textbf{LSTM-AE Testing}\\

Fig.~\ref{fig:training_testing}(b) shows the details of how anomalies can be detected using the reconstruction-based anomaly detection method. Here, the maximum value of MAE from the training process can be designated as the threshold. A reconstruction error rate for each data point from the testing set is compared with this threshold. If the reconstruction loss value is greater than the threshold $\eta$, this data point is noted as an anomaly, otherwise, it is designated to be normal. This is shown in the following Equation \ref{eq:anomalies}.

\begin{equation}\label{eq:anomalies}
X'=
    \left\{ \begin{array}{@{}ll@{}}
       X'_i\ \text{is anomalies}, & \text{if}\ Score > \eta\\
       X'_i\ \text{is normal}, & \text{otherwise}
       \end{array}\right.
\end{equation}
where $X'$ indicates a reconstructed time-series, $X'_i$ is a data point contained in the time-series, and a $score$ is a result of a reconstruction loss function using MAE.


\subsection{Reconstruction-based anomaly detection}
In order to effectively learn time series DDoS traffic behavior using reconstruction-based anomaly detection, our LSTM-AE model is trained with a dataset that contains only benign traffic flows from CICDDoS2019.

\begin{figure*}[t]
	\centering
	\includegraphics[width=0.8\linewidth]{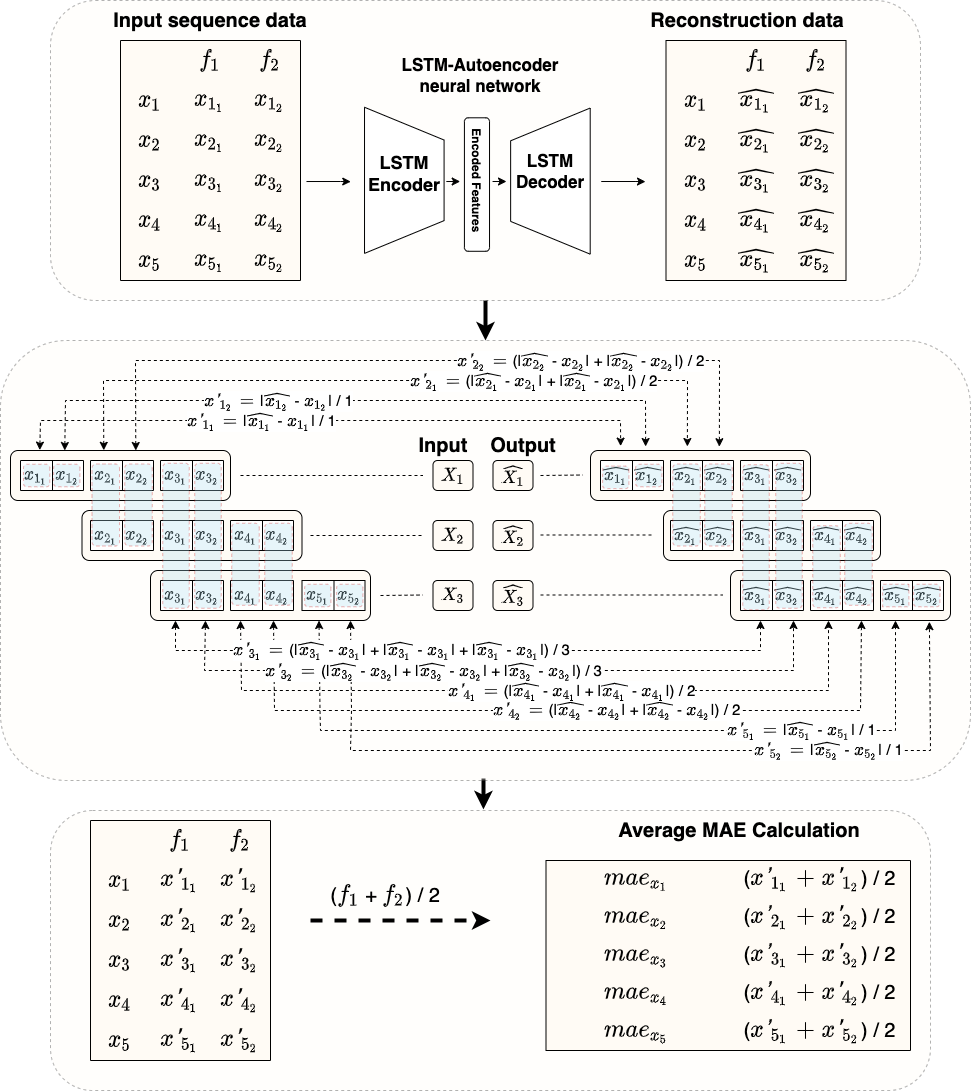}
	\caption{Computing Reconstruction Error on multivariate Time Series} 
	\label{alg:LSTM_AE_anomaly}
\end{figure*}

\subsubsection{LSTM-AE for anomaly detection}
An anomaly can be defined as an observation diverging from the majority of the data. A threshold can be set as a decision point to determine how much an observation deviates. Any observations that exceed the threshold are defined as anomalies. To better demonstrate the functionality of reconstruction-based time series anomaly detection, the LSTM-AE model can be applied to detect anomalies for each input sequence. This allows us to obtain the reconstruction error rates associated with the benign samples of the DDoS dataset. A backpropagation methodology is applied to adjust the weights and parameters of the model. We use the Mean Absolute Error (MAE) algorithm, as shown in Equation~\ref{eq:mae}, as the reconstruction error (loss) function. 
\begin{equation}\label{eq:mae}
	Loss (MAE) = \frac{\sum_{i=1}^{n}\left | x_{i} - \hat{x_i} \right |}{n}
\end{equation}
where $n$ indicates the total number of samples, $x_i$ is the representation of the original input being fed to the encoder, and $\hat{x_i}$ is the output produced by the decoder.

Once training is done and the reconstruction error is computed on all samples, the LSTM-AE model learns a low MAE and sets the maximum reconstruction error as a threshold. By contrast, if the testing set presents different behavior from the training process, the resulting MAE will be greater than the threshold and can be considered an anomaly. To the best of our knowledge of the CICDDoS2019 dataset, there are a few benign samples in each provided CSV file. In order to learn the normal behaviors of traffic flows, we extract all benign samples from the CICDDoS2019 dataset, and designate 80\% of the benign traffic as the training set, while the remaining 20\% of the benign traffic can be combined with different attack types to be used in the testing set.

\subsubsection{Reconstruction Error threshold calculation}
Our proposed reconstruction-based LSTM-AE model was trained in an unsupervised manner and aimed to minimize reconstruction error between input and output while using unlabeled data.  In \cite{wei2022lstm}, the univariate time series LSTM-AE anomaly detection is performed using reconstruction error, shown in Fig.~\ref{alg:LSTM_AE_anomaly}. The figure illustrates how to calculate reconstruction error for each sample contained in the different multivariate time series sequences. Suppose that there is a dataset containing 5 data samples with 2 features (shown as Input sequence data) which are 3 time-series sequences of [$X_1, X_2, X_3$] where each sequence contains 3 samples with 2 features over 3 different timesteps. For example, the first sequence $X_1$ $\in$ \{[$x_{1_1}, x_{1_2}$], [$x_{2_1}, x_{2_2}$], [$x_{3_1}, x_{3_2}$]\}, $X_2$  $\in$  \{[$x_{2_1}, x_{2_2}$], [$x_{3_1}, x_{3_2}$], [$x_{4_1}, x_{4_2}$]\}, and $X_3$ $\in$ \{[$x_{3_1}, x_{3_2}$], [$x_{4_1}, x_{4_2}$], [$x_{5_1}, x_{5_2}$]\}. Our model trains these 3 time series of sequences as inputs and constructs the outputs that map to each sequence $\hat{X_1}$, $\hat{X_2}$, and $\hat{X_3}$.

Using Equation~\ref{eq:mae}, we can obtain the MAE value of each data sample for each feature. In order to obtain the MAE value of each data sample over multivariate features, we use the average MAE of each data sample for each feature. For example, in 3 time-series sequences of [$X_1, X_2, X_3$]. As such, we obtain the reconstruction error of each data sample of each feature. As can be seen in Fig.~\ref{alg:LSTM_AE_anomaly}, with the assumption that each data sample has two features, we can calculate each data sample for each feature using time steps. This figure shows the reconstruction error calculation for each feature when the time step is 3. As an example of one of the data samples of feature 1 ($x_{3_1}$), and the reconstruction error of $x_{3_1}$, the calculation is $(\lvert \hat{x_{3_1}} - x_{3_1}\lvert + \lvert \hat{x_{3_1}} - x_{3_1}\lvert + \lvert \hat{x_{3_1}} - x_{3_1}\lvert) / 3 $, and is defined as $x'_{3_1}$. Similarly, the reconstruction error of feature 2 ($x_{3_2}$) is $(\lvert \hat{x_{3_2}} - x_{3_2}\lvert + \lvert \hat{x_{3_2}} - x_{3_2}\lvert + \lvert \hat{x_{3_2}} - x_{3_2}\lvert) / 3 $, and is defined as $x'_{3_2}$. Note that the reconstructed data sample for each feature can be slightly different, which means each $x_{3_1}$ in $X_1$, $X_2$, and $X_3$ are different. The $x_{3_2}$ is presented with the same definition. After we obtain the reconstruction error of these two features, the total reconstruction error of $x_3$ is calculated by using the average value of features 1 and 2, illustrated as $MAE_{x_3} = (x'_{3_1} + x'_{3_2}) / 2$ where 2 is presented two features of $x_3$. When we obtain all the average reconstruction errors for each data sample on the training set, the maximum average reconstruction error is set as the threshold. During testing, any samples whose reconstruction error goes beyond this maximum average reconstruction error are therefore labeled as anomalies.



\section{Data and Methodologies}
\label{sec:re}

\subsection{CICDDoS2019 Dataset}
In this study, we use the CICDDoS2019~\cite{sharafaldin2019developing} dataset which is widely used for DDoS attack detection and classification. The dataset contains a large number of up-to-date realistic DDoS attack samples as well as benign samples. The total number of records contained in the CICDDoS2019 dataset is depicted in Table~\ref{table:no_dataset}. 

\begin{table} [h]
	\setlength{\tabcolsep}{4.5mm}
	\caption{The number of records in CICDDoS2019 }
	\label{table:no_dataset}
	\begin{tabular}{cccc}
		\midrule 
		{\textbf{dataset}} &
		{\textbf{total}} &
		{\textbf{benign}} & {\textbf{malicious}} \\ 
		\hline
		Training day & 50,063,112	& 56,863	& 50,006,249 \\ 
		Testing day & 20,364,525    & 56,965    & 20,307,560 \\ 
		\midrule
	\end{tabular}
\end{table}

The CICDDoS2019 dataset contains different DDoS attack types that exploit a wide range of network and application protocols. In our study, we used DDoS attack types (i.e., DNS) and benign traffic samples to train and test our proposed LSTM-AE model. The dataset is broken down as follows:

\begin{itemize}
	\item \textbf{Benign}: Benign traffic based on HTTP, HTTPS, FTP, SSH, and email protocols.
	\item \textbf{Attacks}: These DDoS attacks cover two different categories, Reflection-based and Exploitation-based. In terms of our CICDDoS2019 dataset, any traffic included in MSSQL, SSDP, NTP, TFTP, DNS, LDAP, NetBIOS, and SNMP is categorised as a reflection-based attack. Traffic labeled as SYN, UDP, and UDP-lag in CIDDDoS2019 belongs to the exploitation-based category.
\end{itemize}

All CICDDoS2019 data samples from the training day set are depicted in Table~\ref{table:attack_types}. Note that all data samples can be counted from different "CSV" files, which are collected and saved based on their various attack types. Note that the "WebDDoS" attack was collected and saved together with the "UDPLag" attack file.

\begin{table}[h]
	\centering
	\caption{Three Reflection-based Attack Types on Training Day}
	\label{table:attack_types}
	\begin{tabular}{lllll}
	\midrule
	Attack Type & Malicious & Benign & Total Flow count & Attack times \\ \hline
	DNS & 5,071,011 & 3,402 & 5,074,413 & 10:52 - 11:05 \\
    LDAP & 2,179,930 & 1,612 & 2,181,542 & 11:22 - 11:32 \\
    SNMP & 5,159,870 & 1,507 & 5,161,377 & 12:12 - 12:23 \\
    \midrule
    \end{tabular}
\end{table}

The high-level description of the nature of the DDoS attack used in our study is summarised in Table~\ref{table:selected_attacks}. 

\begin{table*}[!h]
	\centering
	\caption{Selected Attack Types in CICDDoS2019 }
	\label{table:selected_attacks}
	\begin{tabular}{ll}
		\midrule
		\textbf{Attack Type} & \textbf{Attack Description} \\
		\midrule 
		DNS Attack & 
		\begin{tabular}[c]{@{}l@{}}DNS attacks are a type of amplification DDoS attack that exploits Domain Name\\ Servers and exhausts the bandwidth of the victims. This attack can overwhelm the victims and make them inaccessible.
		\end{tabular}\\ \hline
		LDAP Attack & 
		\begin{tabular}[c]{@{}l@{}}LDAP attacks are a DDoS attack associated with exploiting Lightweight Directory Access Protocol (LDAP) protocol.\\ The attacker sends a massive number of LDAP requests to the vulnerable LDAP servers by pretending to be a legitimate\\LDAP client using spoofed IP addresses. The LDAP server becomes too busy to create responses for attackers and becomes\\ unable to respond to real LDAP clients.
		\end{tabular}\\ \hline
		SNMP Attack & 
		\begin{tabular}[c]{@{}l@{}}SNMP attacks are a volumetric DDoS attack that stands for Simple Network Management Protocol (SNMP), and these attacks aim \\ to generate a large number of SNMP attacks utilizing a spoofing IP address that directed to the victims from multiple networks.
		\end{tabular}\\ 

		\midrule 
	\end{tabular}
\end{table*}


In this dataset, there are 88 features in total, and the best top 5 features of each attack type and benign traffics have been used based on the RandomForestRegressor class of scikit-learn which can calculate the importance of each feature in the dataset \cite{sharafaldin2019developing}. Table~\ref{table:selected_features} shows the top 5 important features which are employed in this research as well as a brief description.

\begin{table}[h!]
	\centering
	\caption{Selected features in CICDDoS2019}
	\label{table:selected_features}
	\begin{tabular}{ll}
		\midrule 
		Feature Name & Description \\ \hline 
		Max Packet Length & The maximum length of a flow \\ \hline 
		Fwd Packet Length Max & \begin{tabular}[c]{@{}l@{}}The maximum size of packets in the \\ forward direction\end{tabular} \\ \hline
		Fwd Packet Length Min & \begin{tabular}[c]{@{}l@{}}The minimum size of packets in the \\ forward direction\end{tabular} \\ \hline
		Average Packet Size & The average size of packets \\ \hline 
		Min Packet Length & The minimum length of a flow \\
		\midrule 
	\end{tabular}
\end{table}

\subsection{Data Pre-processing}
In this section, we discuss the methodologies we used to process our dataset in order to feed it into our proposed LSTM-AE model.
\subsubsection{Data Cleaning}
The original dataset contained 88 features. As suggested by \cite{sharafaldin2019developing}, they depicted the top 5 most important features of each attack type and benign samples based on weight and mean value calculated. According to these top 5 significant features which were selected, we also use these top-5 features as our features among different attack types. For example, we used DNS, LDAP, and SNMP as our analysis attack types, which provide three attack types (anomalies) as well as benign (normal) samples. We chose these three attack types because they shared the same top 5 important features based on their weight and the mean value calculation. There are five important features to be used in this research: "Max Packet Length", "Fwd Packet Length Max", "Fwd Packet Length Min", "Average Packet Size", and "Min Packet Length". 

\subsubsection{Label Encoding}
We substituted the categorical labels for deep models as they only operate on float/numerical values. One categorical value which was converted was the attack label (benign or an attack type). For label encoding, "0" indicates a benign (normal) sample, while "1" indicates an attack type (anomalies).

\subsubsection{Data Normalization}
There are several widely used methods to perform feature scaling, including Z Score, standardization, and normalization. As proposed by \cite{ur2021diddos}, we utilize the MinMax-based normalization for our feature scaling. This method maps the original range of each feature into a new range using Equation~(\ref{eq:normalised}), as follows:
	
\begin{equation}\label{eq:normalised}
	Z_i = \frac{Z_i - min}{max -  min}
\end{equation}
	
where Z$_i$ denotes all the normalized numeric values ranging between [0-1]; $max$ and $min$ denote the maximum and minimum values from all data points. 

\section{Experimental Results}\label{sec:Experiments}

\begin{table}[h]
	\centering
	\footnotesize
	\caption{Implementation environment specification}
	\label{table:Mat}
	\begin{tabular}{p{2.6cm} | p{3.8cm}}
		\hline
		\textbf{Unit}   & \textbf{Description}\\ \hline
		Processor   & 3.4GH$_z$  Inter Core i5 \\ \hline
		RAM  &  16GB      \\ \hline
		OS  &  MacOS Big Sur   11.4  \\ \hline	
		Packages used  &  tensorflow 2.0.0, sklearn 0.24.1    \\ \hline	
	\end{tabular}
\end{table}

\subsection{Experiment Setup}
Our experiments were carried out using the system setup shown in Table~\ref{table:Mat}. 

The hyperparameters used in the training phase are described (with the values for each parameter) complete with description in Table~\ref{table:Training parameters}.
\begin{table}[h]
	\centering
	\footnotesize
	\caption{LSTM-AE Training Parameters}
	\label{table:Training parameters}
	\begin{tabular}{c|c|c}
		\midrule
		Hyperparameters & Values & Descriptions \\ \midrule
		Learning rate & 0.001 & Learning speed (within range 0.0 and 1.0) \\
		Dropout & 0.2 & No. of neurons ignored \\
		Batch size & 64 & No. of samples in one fwd/bwd pass \\
		Epoch & 30 & No. of one fwd/bwd pass of all samples \\\midrule
	\end{tabular}
\end{table}

\subsection{Performance Matrix}
To evaluate the performance of our model, we used the following metrtics: classification accuracy, precision, recall, and F1 score. Table \ref{table:Matrix} illustrates the confusion matrix.
\begin{table}[h]
	\centering
	\caption{Confusion Matrix}
	\label{table:Matrix}
	\begin{tabular}{| p{2.8cm} | c | p{1.5cm} | p{1.5cm} |}
		\hline
		\multicolumn{2}{|c|}{ \multirow{2}{*}{Total Population} } &   \multicolumn{2}{c|}{Predicted Condition} \\
		\cline{3-4}
		\multicolumn{2}{|c|}{} & Normal & Anomaly \\
		\hline
		\multirow{2}{*}{Actual Condition} & Normal & TN & FP \\
		\cline{2-4}
		&Anomaly & FN & TP \\
		\hline
	\end{tabular}
\end{table}

where;
\begin{itemize}
	\item True Positive (TP) indicates an anomalous data point correctly classified as anomalous. 
	\item True Negative (TN) indicates a normal data point correctly classified as normal.
	\item False Positive (FP) indicates a normal data point incorrectly classified as anomalous.
	\item False Negative (FN) indicates an anomalous data point incorrectly classified as normal.
\end{itemize}

Based on the aforementioned terms, the evaluation metrics are calculated as follows: \\
\begin{equation}\label{eq:TPR}
	TPR (True Positive Rate/Recall) = \frac{TP}{TP + FN}
\end{equation}
\begin{equation}\label{eq:FPR}
	FPR (False Positive Rate) = \frac{FP}{FP + TN}
\end{equation}
\begin{equation}\label{eq:PPV}
	Precision = \frac{TP}{TP + FP}
\end{equation}
\begin{equation}\label{eq:F-measure}
	F1-score = 2\times\left(\frac{Precision\times Recall}{Precision + Recall}\right)
\end{equation}
\begin{equation}\label{eq:ACC}
	Accuracy = \frac{TP+TN}{TP + TN + FP + FN}
\end{equation}
\\
The Area Under the Curve (AUC) computes the area under the Receiver Operating Characteristics (ROC) curve which is plotted using the true positive rate on the y-axis and the false positive rate on the x-axis over different thresholds. Mathematically, the AUC is computed as shown in Equation~(\ref{eq:auc}).

\begin{equation}\label{eq:auc}
	AUC_{ROC}=\int_{0}^{1} \frac{TP}{TP+FN}d\frac{FP}{TN+FP}
\end{equation}


\begin{table*}[!h]
  	\centering
	\footnotesize
	\caption{Three Attack Types' Performance based on different time window lengths}
	\label{table:seq_size}
	\begin{tabular}{l|llll|llll|llll}
     	\midrule
     	\multirow{1}{*}{Attack Type} & \multicolumn{4}{c|}{w = 10ms} & \multicolumn{4}{c|}{w = 50ms} 
     	& \multicolumn{4}{c}{w = 100ms} \\ \hline
     	& Acc & Pre & Re & F1 & Acc & Pre & Re & F1 & Acc & Pre & Re & F1 \\
     	DNS & 96.08 & 99.99 & 94.30 & 97.06 & 95.78 & 99.99 & 93.85 & 96.82 & 95.83 & 99.99 & 93.92 & 96.86 \\
     	LDAP & 99.96 & 99.99 & 99.93 & 99.96 & 98.68 & 99.99 & 97.28 & 98.61 & 99.77 & 99.99 & 97.47 & 98.71 \\
     	SNMP & 96.89 & 99.99 & 95.49 & 97.69 & 96.86 & 99.99 & 95.45 & 97.67 & 96.82 & 99.99 & 95.40 & 97.64 \\
	    \midrule
\end{tabular}
\end{table*}


\begin{table}[h]
	\centering
	\footnotesize
	\caption{Performance Comparison between the experimental results and Three Attack Types on the CICDDoS2019 dataset}
	\label{table:confusion_matrix_table}
    \begin{tabular}{cc|cccc}
    \midrule
    \multicolumn{2}{c|}{\multirow{2}{*}{Algorithms}} & \multicolumn{4}{c}{Confusion Matrix} \\
    \multicolumn{2}{c|}{} & Acc & Pre & Re & F1 \\ \hline
   \addlinespace[0.5ex]
    \multicolumn{2}{c|}{ID3~\cite{sharafaldin2019developing}} & - & 0.78 & 0.65 & 0.69 \\
    \multicolumn{2}{c|}{RF~\cite{sharafaldin2019developing}} & - & 0.77 & 0.56 & 0.62 \\
    \multicolumn{2}{c|}{Naive Bayes~\cite{sharafaldin2019developing}} & - & 0.41 & 0.11 & 0.05 \\
    \multicolumn{2}{c|}{Logistic regression~\cite{sharafaldin2019developing}} & - & 0.25 & 0.02 & 0.04 \\ \hline
    \addlinespace[0.5ex]
    \multirow{3}{*}{Our model} & DNS & 0.96 & 0.99 & 0.94 & 0.97 \\
    & LDAP & 0.99 & 0.99 & 0.99 & 0.99 \\
    & SNMP & 0.96 & 0.99 & 0.95 & 0.97 \\
    \midrule
    \end{tabular}
\end{table}


\begin{figure*}[!htb]
	\centering 
	\begin{subfigure}{0.33\textwidth}
		\includegraphics[width=\linewidth]{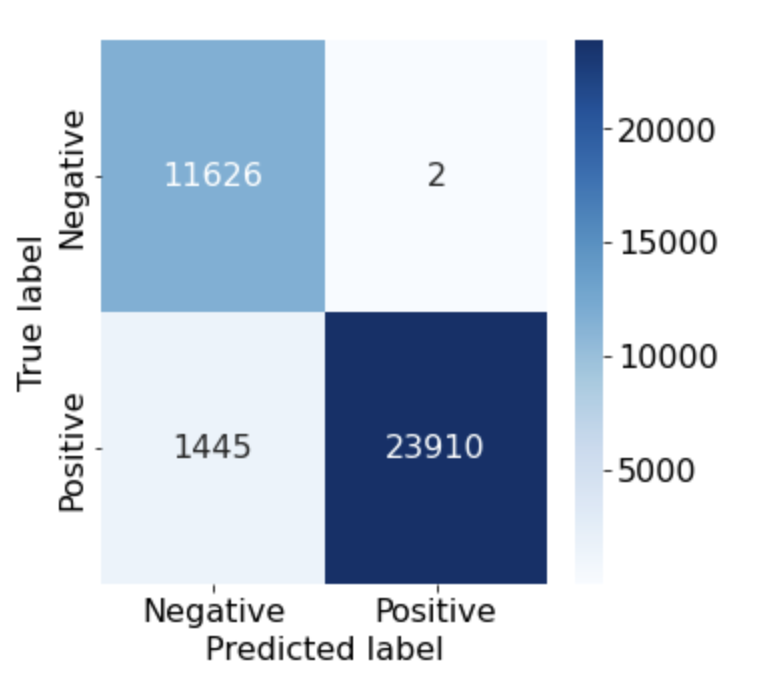}
		\caption{DNS Attack}
		\label{fig:cm_dns}
	\end{subfigure}\hfil 
	\begin{subfigure}{0.33\textwidth}
		\includegraphics[width=\linewidth]{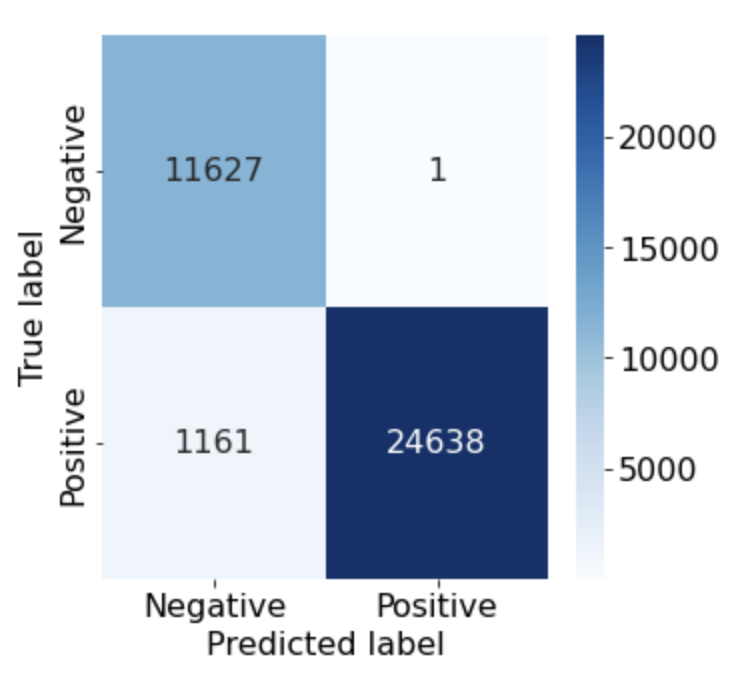}
		\caption{SNMP Attack}
		\label{fig:cm_snmp}
	\end{subfigure}\hfil 
	\begin{subfigure}{0.33\textwidth}
		\includegraphics[width=\linewidth]{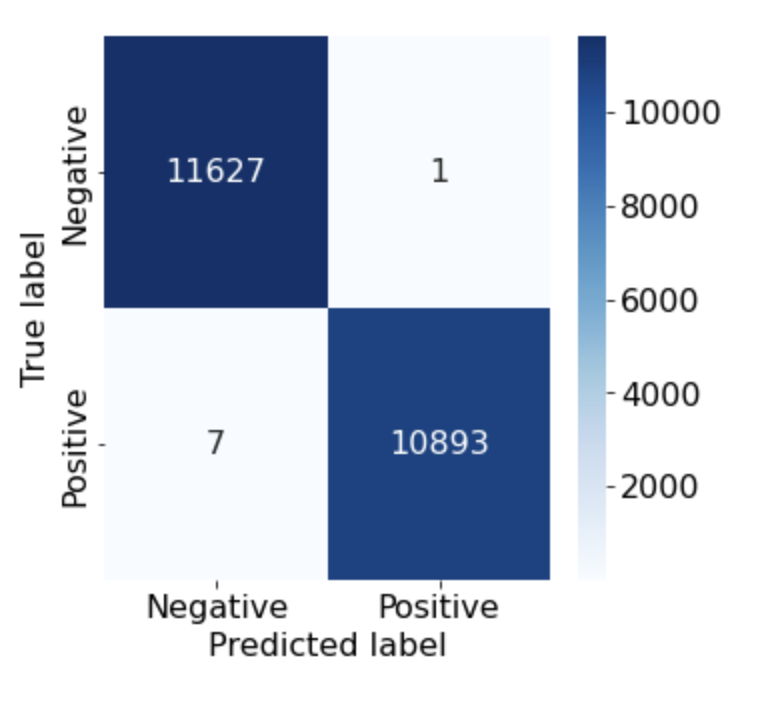}
		\caption{LDAP Attack}
		\label{fig:cm_ldap}
	\end{subfigure}
	\caption{Performance of Anomaly Detection on different DDoS attack types using the Confusion Matrix}
	\label{fig:perf_cm}
\end{figure*}

\subsection{Results and Evaluations}
\subsubsection{Training and testing dataset}
The CICDDoS2019 data collection was based on millisecond intervals. We first select 0.5\% of the data based on the original millisecond time interval. However, after selection, there is a large number of malicious (anomalies) samples with only a few benign samples selected. As mentioned previously, our LSTM-AE model only uses benign samples in the training stage. Thus, to solve this issue, we first extracted all the benign samples from the training day collection and separated them into training (80\%) and testing (20\%) sets. For the model training, we only use 80\% benign samples to train, while the rest of the benign samples can be used in the testing set. Moreover, We used $0.5\%$ of the original CICDDoS2019 dataset for each attack type from the training day collection for testing as it was not feasible to use the full dataset due to performance considerations. Note that each testing set includes benign samples and different attack types.

We experimented with three attack types in the LSTM-AE model to get initial anomaly detection results based on the top 5 most important features for each attack type. Table~\ref{table:confusion_matrix_table} shows the initial anomaly detection results for the three attack types: DNS, LDAP, and SNMP. The results show that the anomaly detection for LDAP attacks performed the best in terms of accuracy, precision, recall, and F1-score, based on the mentioned top 5 features.

Figure~\ref{fig:perf_cm} illustrates the number of records classified for anomaly detection performance using the three DDoS attack types. 

\subsubsection{Influence of time window length}
Table~\ref{table:seq_size} shows the performance of all three attack types based on different time window lengths. The results show that performance was best for detecting LDAP attacks over diverse time window lengths, with over 99\% in accuracy and precision, recall, and F1-score, compared with other attack types as shown in Table~\ref{table:seq_size}. As can be seen in Table~\ref{table:seq_size}, as the time window length increased, the accuracy decreased for all three attack types. The highest accuracy was achieved at over 96\% for a time window length of 10. We also compared the results with the baseline of other machine learning techniques from \cite{sharafaldin2019developing}, in which the same features are used for testing purposes. In the comparison table (Table~\ref{table:confusion_matrix_table}), the evaluation results show our model performed impressively in terms of accuracy, precision, recall, and F1-score.


\subsubsection{Results of batch size}
The testing results were evaluated based on different batch sizes while dropout, learning rate, and epoch remained constant. 

\textit{DNS attack:} For DNS attacks, the highest accuracy was 96.08\% and required a relatively brief computational time (approximately 12s per epoch) with the batch size set to 64. We found that the smallest batch size value tested - size 10 - resulted in the lowest accuracy (88.83\%) and required a relatively high computation time at approximately 42s per epoch.

\begin{table}[h]
	\centering
	\footnotesize
	\caption{Batch Size Performance Metrics for Detecting DNS Attacks}
	\label{table:bs_dns}
	\begin{tabular}{c|cccccc}
        \toprule
		\begin{tabular}[c]{@{}c@{}}Batch \\ size\end{tabular} & Acc & Pre & Re & F1& AUC-ROC & \begin{tabular}[c]{@{}c@{}}Time (s) \\ $(\mu \pm \sigma)$/epoch\end{tabular} \\ \midrule
		\addlinespace[0.5ex]
		10 & 88.83 & 100 & 83.71 & 91.13 & 91.58 & 42 $\pm$ 1.24 \\
		\addlinespace[0.5ex]
		32 & 93.49 & 100 & 90.51 & 95.02 & 95.26 & 12 $\pm$ 1.03 \\
		\addlinespace[0.5ex]
		64 & 96.08 & 99.99 & 94.30 & 97.06 & 97.14 & 12 $\pm$ 0.94 \\
		\bottomrule
	\end{tabular}
\end{table}

\textit{SNMP attack:} In the case of SNMP attacks as in Table~\ref{table:bs_snmp}, analyzing the performances for each batch size demonstrated that the best result was achieved when the batch size was increased to 64. Comparing this with a smaller batch size of 10, all performance metrics showed significant improvement at the larger batch size, particularly computational time. As can be seen in Table~\ref{table:bs_snmp}, a batch size of 10 (38s) takes approximately three times longer than a batch size of 64 (12s), per epoch.

\begin{table}[h]
	\centering
	\footnotesize
	\caption{Batch Size Performance Metrics for Detecting SNMP Attacks}
	\label{table:bs_snmp}
	\begin{tabular}{c|cccccc}
        \toprule
		\begin{tabular}[c]{@{}c@{}}Batch \\ size\end{tabular} & Acc & Pre & Re & F1& AUC-ROC & \begin{tabular}[c]{@{}c@{}}Time (s) \\ $(\mu \pm \sigma)$/epoch\end{tabular} \\ \hline 
		\addlinespace[0.5ex]
		10 & 88.47 & 99.99 & 83.29 & 90.88 & 91.64 & 38 $\pm$ 1.85 \\
		\addlinespace[0.5ex]		
		32 & 96.84 & 99.98 & 95.44 & 97.66 & 97.71 & 14 $\pm$ 1.31 \\
		\addlinespace[0.5ex]
		64 & 96.89 & 99.99 & 95.49 & 97.69 & 97.75 & 12 $\pm$ 0.96 \\
		\bottomrule
	\end{tabular}
\end{table}

\textit{LDAP attack:} The performance of LDAP attacks for each hyperparameter of batch size is presented in Table~\ref{table:bs_ldap}. When analyzing the performance of each batch size, we can see that the LDAP attacks are detected with impressive performance across the different batch sizes, achieving over 98\%. However, a batch size of 10 (42s) takes significantly longer (over three times greater) than a batch size of 10 (12s) per epoch.

\begin{table}[h]
	\centering
	\footnotesize
	\caption{Batch Size Performance Metrics for Detecting LDAP Attacks}
	\label{table:bs_ldap}
	\begin{tabular}{c|cccccc}
        \toprule
		\begin{tabular}[c]{@{}c@{}}Batch \\ size\end{tabular} & Acc & Pre & Re & F1& AUC-ROC & \begin{tabular}[c]{@{}c@{}}Time (s) \\ $(\mu \pm \sigma)$/epoch\end{tabular} \\ \hline 
		\addlinespace[0.5ex]
		10 & 98.82 & 99.99 & 97.57 & 98.76 & 98.78 & 42 $\pm$ 1.93 \\
		\addlinespace[0.5ex]		
		32 & 98.55 & 99.99 & 97.02 & 98.48 & 98.51 & 14 $\pm$ 0.80 \\
		\addlinespace[0.5ex]
		64 & 99.96 & 99.99 & 99.93 & 99.96 & 99.96 & 12 $\pm$ 0.82 \\
		\bottomrule
	\end{tabular}
\end{table}


\subsubsection{Results of learning rate}
The learning rate of the “Adam” optimizer is set at three candidate rates: 0.01, 0.001, and 0.00001. This was tested over the three different attack types, while dropout and epoch values remained constant. 

\textit{DNS attack:} The results of the DNS attacks are shown in Table~\ref{table:lr_dns}, with different learning rates. The highest performance metrics for accuracy, recall, and F1-score was achieved at 96.08\%, 94.30\%, and 97.06\% respectively. These results were all using the learning rate value of 0.001. On the other hand, the smallest learning value at 0.00001 resulted in the lowest accuracy (95.48\%), recall (93.41\%), and F1-score (96.70\%). While selecting different learning rates has an impact on the results, the time taken is not much different per epoch.

\begin{table}[h!]
	\centering
	\footnotesize
	\caption{Learning Rate Performance Metrics for Detecting DNS Attacks}
	\label{table:lr_dns}
	\begin{tabular}{c|cccccc}
        \toprule
		\begin{tabular}[c]{@{}c@{}}Learning \\ Rate\end{tabular} & Acc & Pre & Re & F1& AUC-ROC & \begin{tabular}[c]{@{}c@{}}Time (s) \\ $(\mu \pm \sigma)$/epoch\end{tabular} \\ \hline 
		\addlinespace[0.5ex]
		0.001 & 96.08 & 99.99 & 94.30 & 97.06 & 97.14 & 10 $\pm$ 0.99 \\
		\addlinespace[0.5ex]		
		0.0001 & 95.80 & 99.99 & 93.88 & 96.84 & 96.94 & 10 $\pm$ 0.84 \\
		\addlinespace[0.5ex]
		0.00001 & 95.48 & 99.99 & 93.41 & 96.59 & 96.70 & 10 $\pm$ 1.05 \\
		\bottomrule
	\end{tabular}
\end{table}

\textit{SNMP attack:} In examining the performance of detection on the SNMP attack in Table~\ref{table:lr_snmp}, we can see that the best performance for accuracy was achieved at 96.89\% using the learning rate of 0.001. Moreover, the results of all performance metrics decrease slightly as the learning rate changed from 0.001 to 0.00001. The processing time did not vary significantly due to the changes in the learning rate.

\begin{table}[h!]
	\centering
	\footnotesize
	\caption{Learning Rate Performance Metrics for Detecting SNMP Attacks}
	\label{table:lr_snmp}
	\begin{tabular}{c|cccccc}
        \toprule
		\begin{tabular}[c]{@{}c@{}}Learning \\ Rate\end{tabular} & Acc & Pre & Re & F1& AUC-ROC & \begin{tabular}[c]{@{}c@{}}Time (s) \\ $(\mu \pm \sigma)$/epoch\end{tabular} \\ \hline 
		\addlinespace[0.5ex]
		0.001 & 96.89 & 99.99 & 95.49 & 97.69 & 97.75 & 12 $\pm$ 0.96 \\
		\addlinespace[0.5ex]		
		0.0001 & 96.82 & 99.99 & 95.39 & 97.63 & 97.69 & 12 $\pm$ 0.99 \\
		\addlinespace[0.5ex]
		0.00001 & 96.88 & 99.99 & 95.48 & 97.68 & 97.74 & 12 $\pm$ 1.05 \\
		\bottomrule
	\end{tabular}
\end{table}

\textit{LDAP attack:} As shown in Table~\ref{table:lr_ldap}, the LDAP attack performed the best over our experiments when compared to the other attack types. The results of detection overall metrics - accuracy, precision, recall, F1, and AUC-ROC - provided the smallest variation over LDAP attacks using the different learning rates, all scores being over 99\%. The processing times showed there was a slightly greater cost using the smallest learning rate of 0.00001.

The impact on the different learning rate configurations presents little variation, but a learning rate of 0.001 gave the best results for accuracy. Similarly, the lowest value for the learning rate took a long time to converge, which results in a significantly longer time for computation per epoch. 

\begin{table}[h!]
	\centering
	\footnotesize
	\caption{Learning Rate Performance Metrics for Detecting LDAP Attacks}
	\label{table:lr_ldap}
	\begin{tabular}{c|cccccc}
        \toprule
		\begin{tabular}[c]{@{}c@{}}Learning \\ Rate\end{tabular} & Acc & Pre & Re & F1& AUC-ROC & \begin{tabular}[c]{@{}c@{}}Time (s) \\ $(\mu \pm \sigma)$/epoch\end{tabular} \\ \hline 
		\addlinespace[0.5ex]
		0.001 & 99.96 & 99.99 & 99.93 & 99.96 & 99.96 & 11 $\pm$ 0.91 \\
		\addlinespace[0.5ex]		
		0.0001 & 99.94 & 99.99 & 99.89 & 98.94 & 99.95 & 11 $\pm$ 0.90 \\
		\addlinespace[0.5ex]
		0.00001 & 99.96 & 99.99 & 99.93 & 99.96 & 99.96 & 12 $\pm$ 0.94 \\
		\bottomrule
	\end{tabular}
\end{table}

\begin{figure*} [!h]
  \centering
  \begin{subfigure}[b]{0.3\textwidth}
   \begin{adjustbox}{width=1.05\linewidth} 
	\begin{tikzpicture}
		\begin{axis}[
			font=\sf,
			xlabel={False Positive Rate},
			ylabel={True Positive Rate},
			xmin= 0, xmax=1,
			ymin= 0, ymax=1,
			xtick={0,.2,.4,.6,.8,1},
			ytick={0,.2,.4,.6,.8,1},
			legend pos=south east,
			legend image post style={scale=0.2},
			no markers,
			]
			\addlegendimage{empty legend}
			\addlegendentry{\textbf{AUC}}
			
			\addplot[smooth,red, tension=0.2] 
			coordinates {(0,0)(0.00,0.9430)(1,1)}; 
			\addlegendentry{DNS =  97.14}
			
			\addplot[smooth,green, tension=0.2] 
			coordinates {(0,0)(0.00,0.9549)(1,1)}; 
			\addlegendentry{SNMP =  97.75}
			
			\addplot[smooth,purple, tension=0.2] 
			coordinates {(0,0)(0.00,0.9993)(1,1)}; 
			\addlegendentry{LDAP =  99.96}
			
			\addplot[black,dashed] coordinates{(0,0) (1,1)};
		\end{axis}
	\end{tikzpicture}
	\end{adjustbox}
	\caption{time window length: 10}
	\end{subfigure}
	\hfill
  \begin{subfigure}[b]{0.3\textwidth}
   \begin{adjustbox}{width=1.05\linewidth}
	\begin{tikzpicture}
		\begin{axis}[
			font=\sf,
			xlabel={False Positive Rate},
			ylabel={True Positive Rate},
			xmin= 0, xmax=1,
			ymin= 0, ymax=1,
			xtick={0,.2,.4,.6,.8,1},
			ytick={0,.2,.4,.6,.8,1},
			legend pos=south east,
			legend image post style={scale=0.2},
			no markers,
			]
			\addlegendimage{empty legend}
			\addlegendentry{\textbf{AUC}}
			
			\addplot[smooth,red, tension=0.2] 
			coordinates {(0,0)(0.00, 0.9385)(1,1)}; 
			\addlegendentry{DNS =  96.92}
			
			\addplot[smooth,green, tension=0.2] 
			coordinates {(0,0)(0.00, 0.9545)(1,1)}; 
			\addlegendentry{SNMP =  97.72}
			
			\addplot[smooth,purple, tension=0.2] 
			coordinates {(0,0)(0.00, 0.9728)(1,1)}; 
			\addlegendentry{LDAP =  98.64}
			
			\addplot[black,dashed] coordinates{(0,0) (1,1)};
		\end{axis}
	\end{tikzpicture}
	\end{adjustbox}
	\caption{time window length: 50}
	\end{subfigure}
	\hfill
  \begin{subfigure}[b]{0.3\textwidth}
   \begin{adjustbox}{width=1.05\linewidth}
	\begin{tikzpicture}
		\begin{axis}[
			font=\sf,
			xlabel={False Positive Rate},
			ylabel={True Positive Rate},
			xmin= 0, xmax=1,
			ymin= 0, ymax=1,
			xtick={0,.2,.4,.6,.8,1},
			ytick={0,.2,.4,.6,.8,1},
			legend pos=south east,
			legend image post style={scale=0.2},
			no markers,
			]
			\addlegendimage{empty legend}
			\addlegendentry{\textbf{AUC}}
			
			\addplot[smooth,red, tension=0.2] 
			coordinates {(0,0)(0.00,0.9392)(1,1)}; 
			\addlegendentry{DNS =  96.96}
			
			\addplot[smooth,green, tension=0.2] 
			coordinates {(0,0)(0.00,0.9540)(1,1)}; 
			\addlegendentry{SNMP =  97.70}
			
			\addplot[smooth,purple, tension=0.2] 
			coordinates {(0,0)(0.00,0.9747)(1,1)}; 
			\addlegendentry{LDAP =  98.73}
			
			\addplot[black,dashed] coordinates{(0,0) (1,1)};
		\end{axis}
	\end{tikzpicture}
	\end{adjustbox}
	\caption{time window length: 100}
  \end{subfigure}
  \caption{AUC-ROC Visualization}
  \label{fig:AUC_ROC}
\end{figure*}
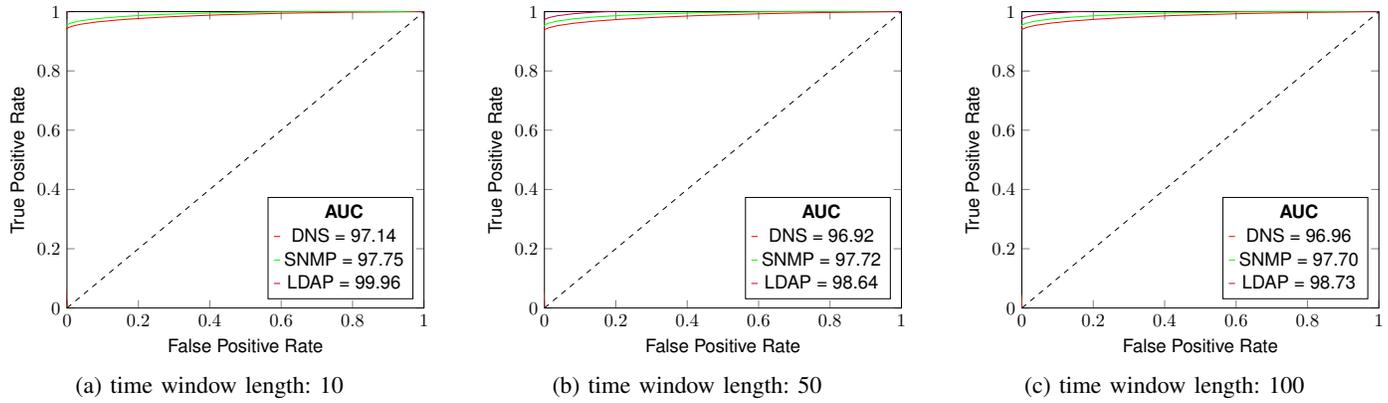

Figure~\ref{fig:AUC_ROC} shows the AUC-ROC for anomaly detection over our three DDoS attack types using different time window length selections. Note that all three attack types achieved over 97\% for the AUC, and that the highest value for the AUC-ROC was achieved for a time window length of 10. Our experimental evidence shows that the proposed model offers significant potential to detect DDoS attacks effectively.


Table~\ref{table:Comparison} shows the performance comparison for our proposed LSTM-AE model with other similar methods from shallow machine learning and deep learning-based approaches. As the results demonstrate, our approach offered the best performance of LDAP in terms of all aspects of evaluation metrics, reaching an average of 99.96\% accuracy while precision, recall, and F1-score all remain very competitive at 99.99\%, 99.93\%, and 99.96\% respectively.

\begin{table*}[h]
    \centering
	\caption{Comparison of CICDDoS2019 with Different Algorithms}
	\label{table:Comparison}
	\begin{tabular}{c|c|c|c|c|c}
	\midrule
	Paper      & Techniques       & Accuracy & Precision & Recall  & F1-score \\ \hline
	Novaes et al.~\cite{novaes2020long}        & LSTM-Fuzzy              &  -  &     & 93.13  &    \\ \hline
	\multirow{2}{*}{Can et al.~\cite{can2021detection}} & DDoSNet 24 features                                                & - & 91.12 & 72.91 & 74.00 \\ 
	& \begin{tabular}[c]{@{}l@{}}DDosNet 82\\ features + FS\end{tabular} & - & 91.16 & 79.41 & 79.39 \\ \hline
	Elsayed et al.~\cite{elsayed2020ddosnet}  & DDosNet (RNN+AE) & 99       & 99        & 99      & 99       \\ \hline
	liao et al.~\cite{liao2022traffic} & K-means + Active Learning Method & 90 & - & - & 95 \\ \hline
	Alghazzawi et al.~\cite{alghazzawi2021efficient} & CNN + BiLSTM & 94.52 & 94.74 & 92.04 & 93.44 \\ \hline
	Jia et al.~\cite{jia2020flowguard} & LSTM & 98.9 & 99.47 & 99.31 & 99.35 \\ \hline
	Gniewkowski et al.~\cite{gniewkowski2022anomaly} & LSTMED & 89.7 & 44.8 & 88.3 & 59.4 \\ \hline	Sayed et al.~\cite{sayed2022multi} & LSTM(scenario II) & 88.5 & 88.1 & 87.8 & 87.0 \\ \hline
	This paper & LSTM-AE           &  99.96   &  99.99    &  99.93  &  99.96  \\ \midrule
	\end{tabular}
\end{table*}

\section{Conclusion}\label{sec:Conclusion}
In this study, we demonstrate that DDoS attacks can be detected with high accuracy using a combination of multiple deep learning-based techniques. Our proposed reconstruction-based LSTM-AE anomaly detection model leverages the benefits of an LSTM model and an Autoencoder in order to detect DDoS traffic anomalies. We use LSTM networks consisting of multiple LSTM units that work with each other to learn the long short-term correlation of DDoS traffic within a time series sequence. An Autoencoder is employed to identify the optimal threshold based on the reconstruction error rates. This can be used to identify anomalies in traffic. We have demonstrated the impact of different window lengths for classifying anomalies over different DDoS attack types. Our proposed model offers potential as an effective DDoS defense tool to assist in detecting a massively growing number of DDoS anomalies. Our model has been comprehensively and extensively tested against three different DDoS attack types. The evaluation results demonstrate high levels of performance on different time window lengths over many performance metrics including precision (99\%), recall (99\%), F1-score (99\%), and accuracy (99\%). Our model performed best for the LDAP attack detection case against all performance metrics, exceeding 99\% and outperforming other state-of-the-art methods.

\bibliographystyle{IEEEtran}
\bibliography{mybib}

\end{document}